\documentclass[twoside,twocolumn,fleqn,10pt]{elsart}
\usepackage{amsmath,epsfig,psfrag,graphicx,subfigure}
\usepackage{fancyheadings}
\def\etal{{\sl et al.}}

\def\pic{PIC}                   
\def\fc{FC}                     
\def\bc{BSC}                     
\def\beamchamber{beam strip chamber} 

\begin{document}
\begin{frontmatter}

\title{Ionization Chambers for Monitoring in High-Intensity
  Charged Particle Beams}

\author[Pitt]{J.~McDonald\corauthref{cor}}, 
\corauth[cor]{Corresponding author.}
\ead{jemcdon@pitt.edu}
\author[Wisc]{C.~Velissaris}, 
\author[BNL]{B.~Viren}, 
\author[BNL]{M.~Diwan}, 
\author[Wisc]{A.~Erwin}, 
\author[Pitt]{D.~Naples} and
\author[Wisc]{H.~Ping}

\address[Pitt]{Department of Physics and Astronomy,
    University of Pittsburgh, Pittsburgh, PA 15260}
\address[Wisc]{Department of Physics, University of Wisconsin--
    Madison, Madison, WI, 53706}
\address[BNL]{Brookhaven National Laboratory, Upton, NY
    11973-5000}

\begin{abstract} 
  Radiation-hard ionization chambers were tested using an intense
  electron beam from the accelerator test facility (ATF) at the
  Brookhaven National Laboratory (BNL).  The detectors were designed
  to be used as the basic element for monitoring muons in the Main
  Injector Neutrino beamline (NuMI) at the Fermi National Accelerator
  Laboratory (FNAL).  Measurements of linearity of response, voltage
  dependence, and the onset of ionization saturation as a function of
  gap voltage were performed.
\end{abstract} 

\end{frontmatter}

\section{Introduction}
\label{intro}

Future neutrino experiments will drive neutrino beams to intensity
levels where beam secondary and tertiary monitoring systems must be
concerned about effects of radiation damage on materials and detector
response linearity.  We have designed a radiation-hard ceramic pad
ionization chamber (PIC) with small (adjustable 1-5~mm) gap
thickness suitable for this environment.~\cite{NuMiTRD} The device
will be used to monitor secondary hadrons and muons in the NuMI
neutrino beam for the MINOS experiment.  This paper presents
measurements performed at the ATF which demonstrate that the system
can be used for monitoring charged particle intensities in the range
$10^6-10^8$ particles/cm$^2$ in an 8$\mu$sec beam spill, suitable for
which are the expected intensities at the MINOS muon monitors.

\section{Minos Beam Environment}

The NuMI beam, composed primarily of $\nu_{\mu}$, will be produced
from decays of secondaries generated when protons from the Main
Injector at FNAL strike a graphite target.~\cite{NuMiTRD} The
neutrinos emerge from \mbox{$\pi^{+}\rightarrow \mu^{+}\nu_{\mu}$} and
\mbox{$K^{+}\rightarrow \mu^{+}\nu_{\mu}$} decays in a 675~m decay
pipe beginning 50~m downstream of a double horn focusing system.

\begin{figure} 
\begin{center}
  \includegraphics[width=0.97\linewidth]{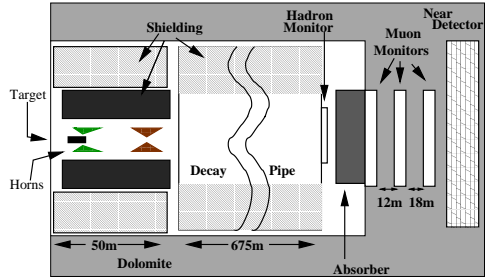}
\end{center}
\caption{Schematic of the NuMI neutrino beamline.  The absorber region
  and the locations of the beam monitoring stations are indicated in
  the right hand side of the figure.  Figure is not to scale.}
\label{mualcoves} 
\end{figure}

In order to predict the beam flux at the far detector MINOS plans to
use the neutrino beam spectrum measured at the near detector.
However, neutrino events are collected too slowly at the near detector
for spill by spill monitoring of the beamline performance.  Beam
monitors will be used to measure the distributions of secondary
particles at the end of the decay pipe and of muons produced in the
$K$ and $\pi$ decays at three locations (muon alcoves) in the earth,
see Fig.~\ref{mualcoves}.  This system is designed to be sensitive to
beamline component alignment and performance and to provide feedback
on the beam on a short time scale.\cite{NuMiTRD}

In order to monitor the muons in the beam, arrays of PIC detectors
will be located immediately downstream of the hadron absorber and
successively after 12~m and 18~m of earth in the muon alcoves.  To
ensure that these chambers will operate in the linear region, we have
measured the saturation point (defined later in
Sec.~\ref{Results}), the ionization response yield, the stability
of the yield and its dependence on the applied electric field, as
described in this paper.

\section{ATF Beam}

The Accelerator Test Facility (ATF) at Brookhaven National Laboratory
provided the test beam to study the PIC response.  The facility
delivered 42 MeV electron bunches with pulse lengths of $\sim$10~ps
and a range of intensities from 1~pC to 1~nC.~\cite{atf} The beam spot
size was typically between 1.0 and 1.2~cm$^2$ (85\% containment), thus
the typical intensity range available at the facility was
$6\times10^6-6\times10^{9}$ particles/cm$^2$/10~ps.

\section{PIC Description}
\label{description}

\begin{figure} 
\begin{center}
  \includegraphics[width=0.70\linewidth]{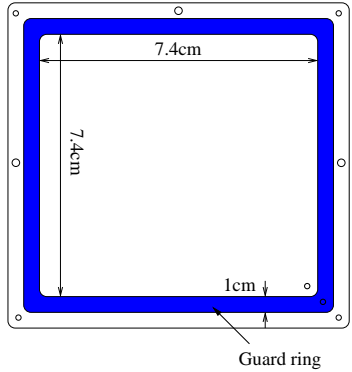}
\end{center}
\caption{Front view of \pic{} signal pad.}
\label{fig:front} 
\end{figure}

The \pic{} is made of two metalized ceramic pads separated by laser
cut ceramic washers at the corners to maintain the gap spacing.  The
ceramic pads are approximately 7.4~cm $\times$ 7.4~cm, a signal pad is
shown in Fig.~\ref{fig:front}.  A 0.304 mm gap separates the 1~cm
thick guard ring from the active pad.  The metalized-ceramic pad is
produced commercially 
using a Platinum-Silver alloy and 1~mm thick ceramic wafers as the
substrate by Amitron, Inc. of N.  Andover, MA, USA.  The components of
the basic \pic{} unit have been chosen for their radiation hardness
and flatness of less than 1.35 $\mu$m per cm.  The radiation dose in
the muon alcoves will be approximately 10~MRads/year.~\cite{Harris}
Mechanical Properties of ceramic materials have been tested up to
total radiation doses of 10~GRad.~\cite{CERN8912}

\section{Experimental Setup}
\label{setup}

\begin{figure} 
\centering
\mbox{\subfigure[]{\includegraphics[width=0.40\linewidth]{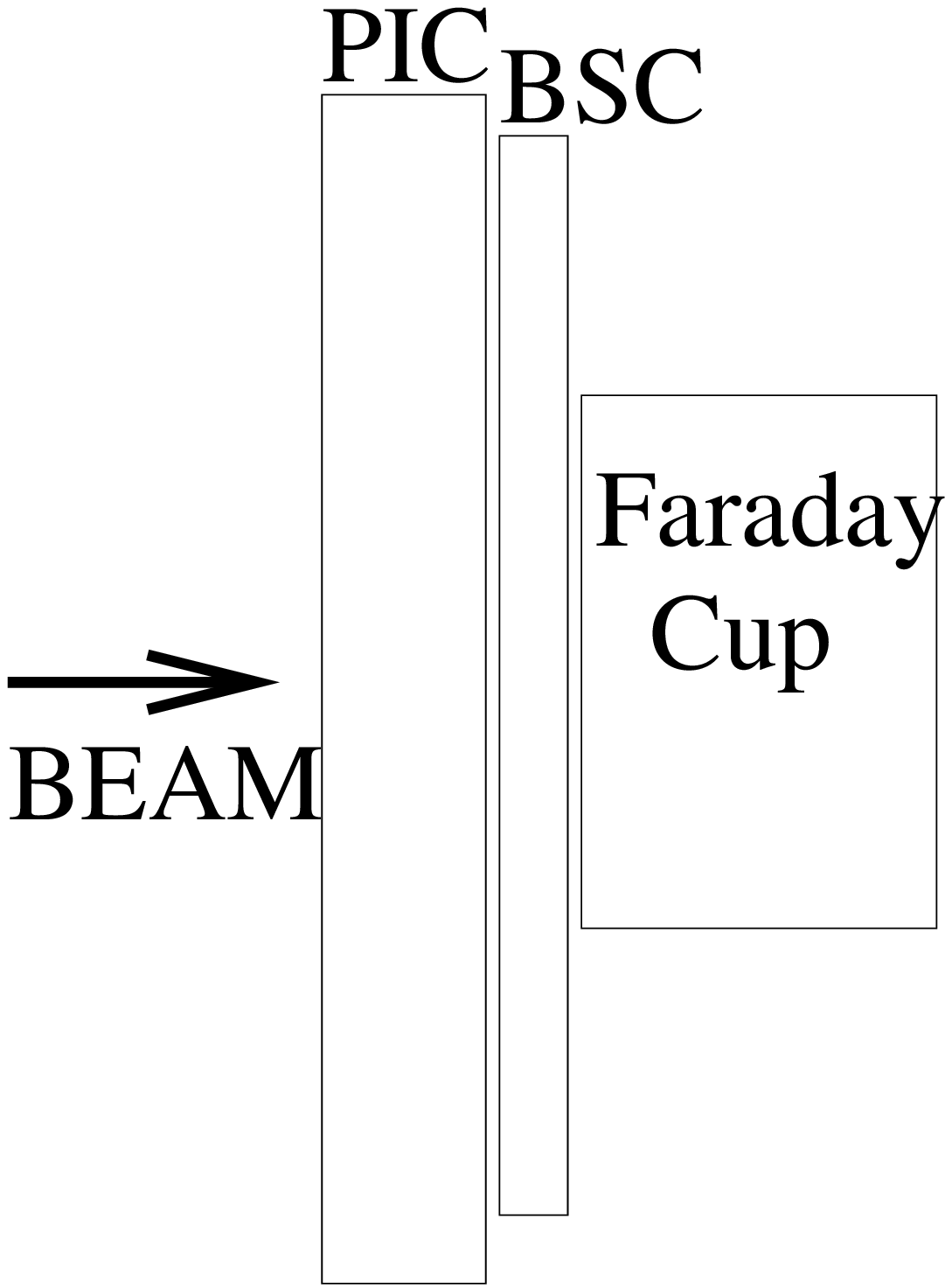}}
\hspace{0.05\linewidth}\subfigure[]{
\includegraphics[width=0.40\linewidth]{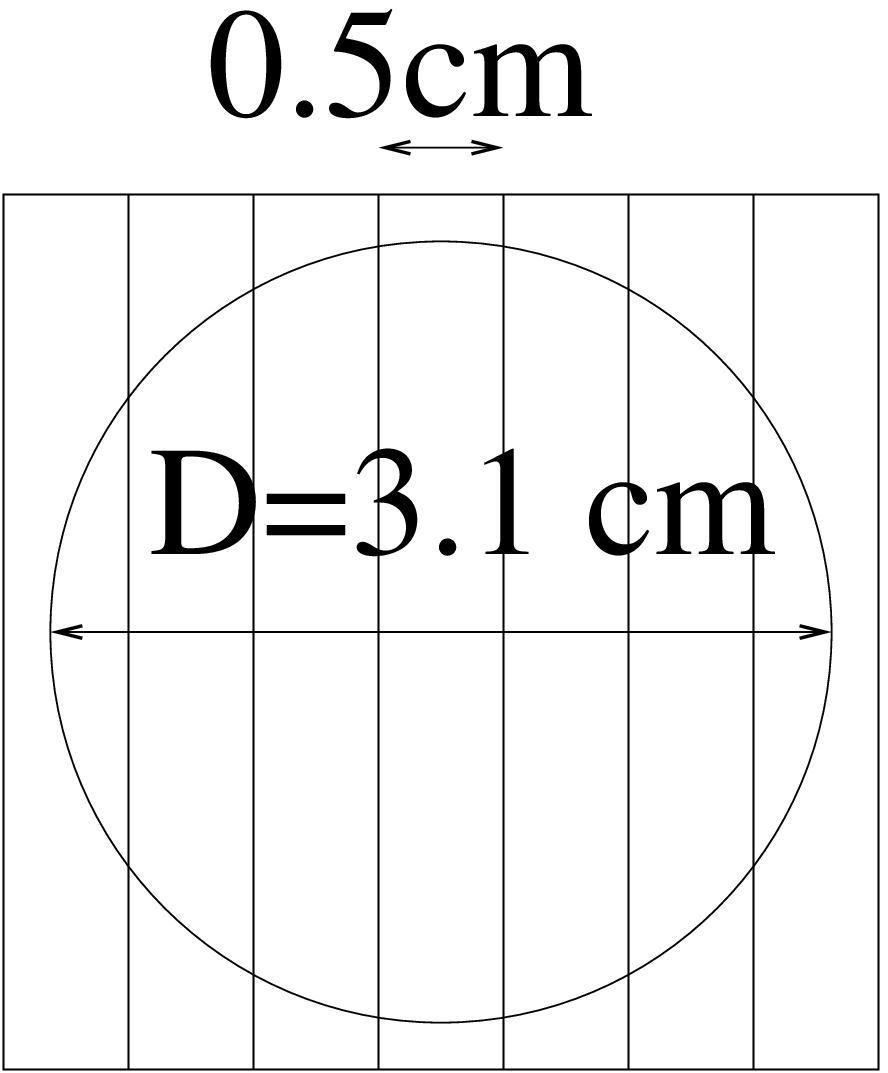}}}
\caption{Schematic of the apparatus showing (a) the detector locations and (b) the front view of the \beamchamber{} shadowed by the Faraday cup showing the active area of each.}
\label{fig:bc} 
\end{figure}

The experimental setup is shown in Fig.~\ref{fig:bc}a.  The \pic{}
detector pads were mounted in a thin window gas box upstream of the
Faraday cup (\fc{}) and the \beamchamber{} (\bc{}). The \fc{} measured
the total beam charge per pulse.  The \bc{} was used to measure the
beam profile and consisted of a pair of strip ionization chambers.
Each chamber had a set of seven 5~mm wide strips: one set oriented
horizontally and the other vertically.  Each had a gap size of 2.5~mm
per plane.  Both the \bc{} and the \fc{} completely covered the
expected beam size of \mbox{$<$ 1.5~cm$^2$} (see Fig.~\ref{fig:bc}b).

The \bc{} and the \pic{} detectors were operated with helium flowing
at a rate of 100 $\pm$ 20 cc/min and at a pressure of 0.15 $\pm$ 0.05
inches of water above ambient pressure.  The \fc{} was also operated
in a continuously purged atmosphere of dry helium.  We used laboratory
grade (UHP) helium independently rated to be 99.999\% pure by Praxair
Inc., Seattle, WA, USA.
The helium gas was provided separately to the \pic{}, \bc{} and the
plastic enclosure sealed with kapton tape around the entire assembly.
A gas distribution manifold allowed us to control the flow rate and
pressure to each detector.  We used 0.25~{\it in} polyethylene tubing
with brass fittings to distribute the gas.  No provision was made to
sample the gas for purity.

Voltage to the detectors was provided by two independent high voltage
units with local filters to suppress ripples.  For all of the data,
the \bc{} voltage remained fixed at 250~V.

\subsection{Electronics}
\begin{figure} 
\begin{center}
\includegraphics[width=0.80\linewidth]{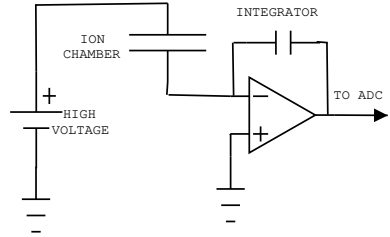} 
\end{center}
\caption{Schematic of the ionization chamber and input stage
  of the electronics.}
\label{fig:setup} 
\end{figure}

The essential features of the electronic system used to acquire data
are shown in the simplified schematic of Fig~\ref{fig:setup}.  An
operational amplifier removed excess charge from either an ionization
chamber with capacitance 18~pF or the Faraday cup and the connecting
coax cable, and stored it on a precision 100~pF integration capacitor.
The voltage on the capacitor was subsequently digitized by an ADC for
computer read out.  Voltage on the input to the integrator was
maintained at a virtual ground subject to the limitations of the
amplifier. This differs from pulse height read out systems that employ
a load resistor to develop a voltage at this point in the circuit.

The operational amplifier is part of a Burr-Brown ACF2101 integrated
circuit.~\cite{burrbrown} The complete integrated circuit contains the
necessary FET switches for clearing the integration capacitor and
gating input/output signals. Notable limitations of the amplifier are
a 3~V/$\mu$sec slew rate and a random 1 mV charge transfer noise at
the output.

Charge integration electronics distinguish these ionization chamber
measurements from previous accelerator measurements which used pulse
height read out.~\cite{Palestini:1998an,boag} Pulse height is more
susceptible to saturation effects because pulse height and shape
depend on charge collection time as well as the amount of charge
produced in the detector.

The complete system used for data acquisition was developed by FNAL
for read out of a segmented wire ionization chamber
(SWIC).~\cite{swic} The system used here differs from the FNAL system
by the addition of a bus termination resistor to reduce channel cross
talk for large output signals.

A short integration gate time of 5~msec for readout was necessary to
prevent a small charge loss that was observed for much longer gate
times.

In the SWIC electronics, a large contribution to the error comes from
random charge transfer errors of 0.1~pC.  This contributes
significantly to the measurement uncertainty at \fc{} charges of
$\approx$ 1~pC.  All pedestals were taken while the beam was disabled.
The pedestals did not vary with the \pic{} operating voltage.  The
pedestals for the \pic{} and \fc{} show stability over time and show
no signs of systematic deviation which would be manifest if, for
example, a dark current were present.  For the PIC detector, a typical
pedestal was measured to be \mbox{0.39 pC} with a typical width of
\mbox{0.13 pC}.  The pedestal width includes both statistical and the
dominant charge transfer error.  

\section{Analysis}
\label{analysis}

We recorded the integrated charge for the \pic{}, all channels of the
\bc{} and the \fc{} for each ATF pulse.  Data were pedestal subtracted
and then selected for beam containment.  The ratio of the the \pic{}
to \fc{} charge in the linear region was examined and anomalous events
were removed.

The mean horizontal beam position was determined from the \bc{} using
\(X_c =\Sigma q_i i/\Sigma q_i\) where $i$ is the strip number and
$q_i$ is the charge collected on strip $i$.  $Y_c$, the vertical
position was determined similarly.  The horizontal and vertical beam
widths were estimated using the standard deviation (RMS) of the beam
intensity in each view.  For the horizontal, this width is
\(\sigma_{X_c}=\sqrt{\frac{\Sigma q_i i^2}{\Sigma q_i} -
  \left(\frac{\Sigma q_i i}{\Sigma q_i}\right)^2}\).  It is similarly
defined for the vertical width.

The beam position was stable for all charges and the beam width was
usually less than one strip.  Cuts were made on the beam position and
horizontal and vertical widths to select well behaved beam pulses.
During a typical run, the efficiency of the cuts was 99\%.

\section{Results}
\label{Results}

In gas ionization detectors, particles ionize the atoms of the gas
generating electrons and ions which in turn generate a detectable
current if an external electric field is applied.  With a suitable
electric field, the detectors operate in ionization mode where the
number of ionized gas particles is simply proportional to the total
number of incident charged particles.  If the applied field is too
large, the ionized charges are accelerated enough to produce
additional ionization, and gas amplification occurs.

Saturation occurs when the applied field is not large enough to
overcome the field produced by charges from ionization. In this case,
ionized charges can recombine and any further increase in beam
intensity does not yield a corresponding increase in ionization.  At
voltages below the gas amplification region, saturation can be
overcome by increasing the external electric field or decreasing the
amount of material in the active ionization region so that less gas is
ionized.  The literature on saturation does not provide a complete
physical description under all experimental
conditions.~\cite{Palestini:1998an,boag} The saturation point can be
experimentally determined by measuring the deviation of the response
from linear or by measuring the ionization response yield, i.e.
plateau curves as a function of voltage.

\subsection{Linearity}

\begin{figure} 
\begin{center}
\includegraphics[width=0.70\linewidth,angle=270]{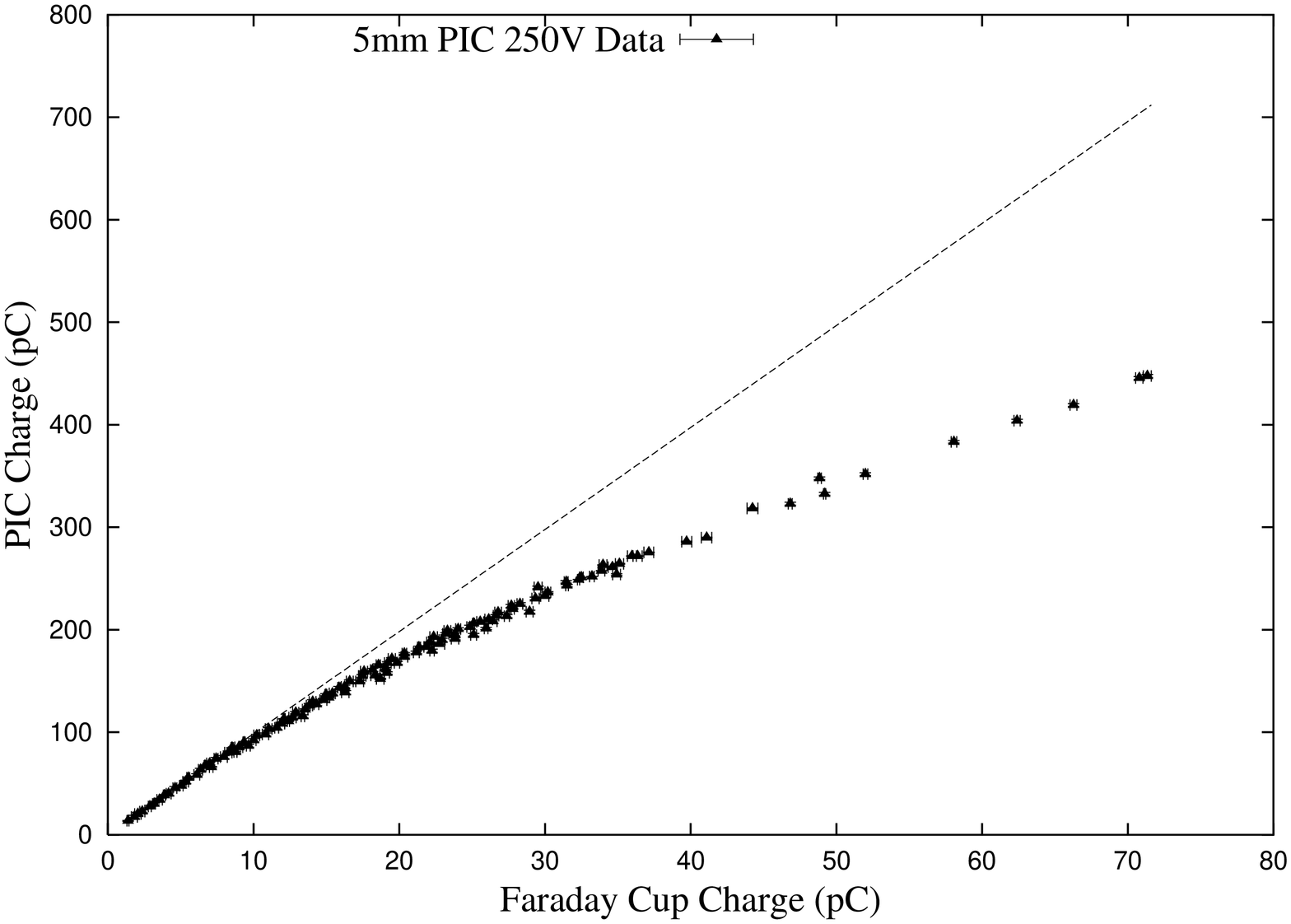}
\end{center}
\caption{Response of the \pic{} versus the \fc{} charge for 5~mm 250~V data.}
\label{fig:intscan5mm} 
\end{figure}

\begin{figure} 
\begin{center}
\includegraphics[width=0.70\linewidth,angle=270]{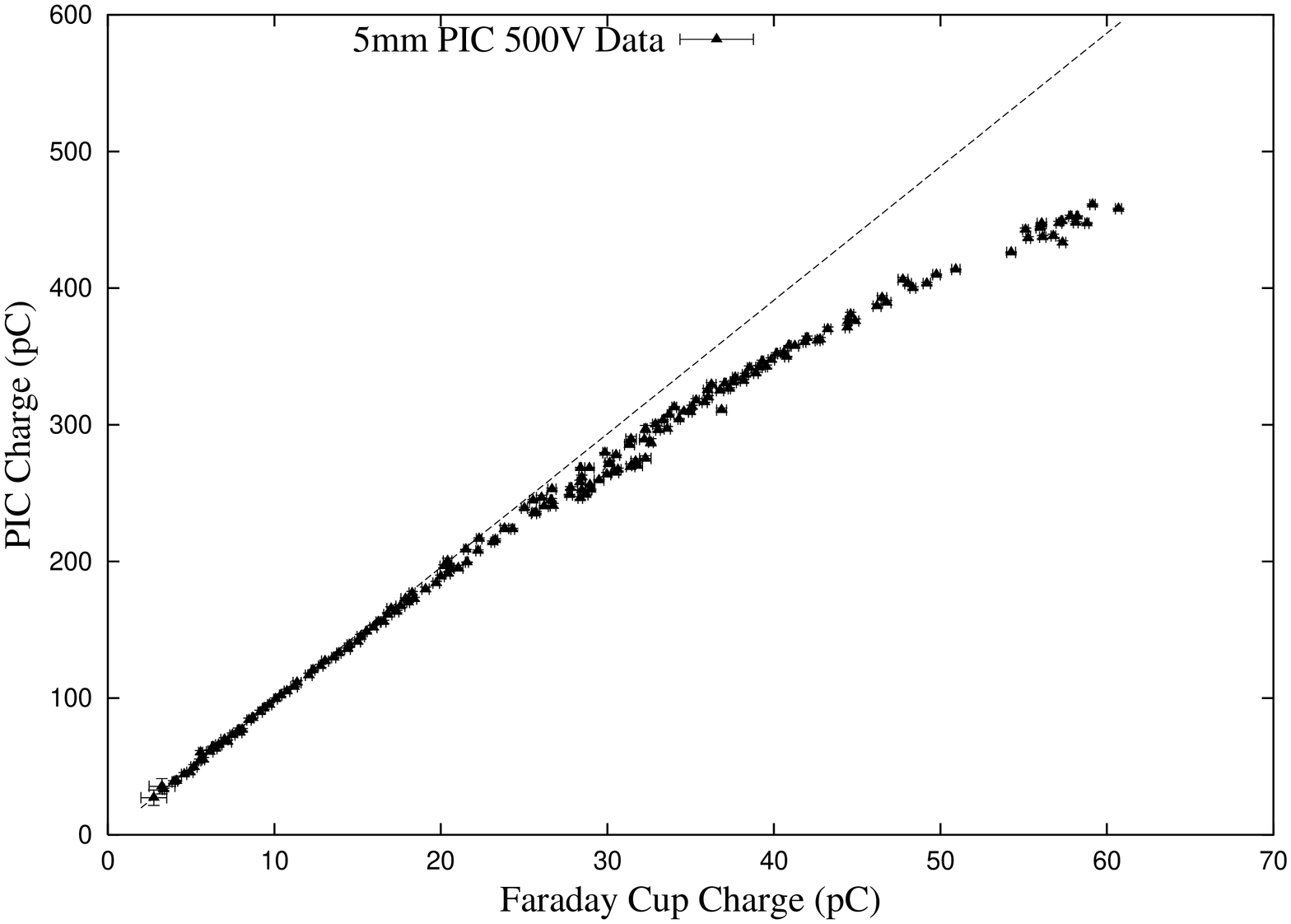}
\end{center}
\caption{Response of the \pic{} versus the \fc{} Charge for 5~mm 500~V data.}
\label{fig:intscan5mm2} 
\end{figure}

\begin{figure} 
\begin{center}
\includegraphics[width=0.70\linewidth,angle=270]{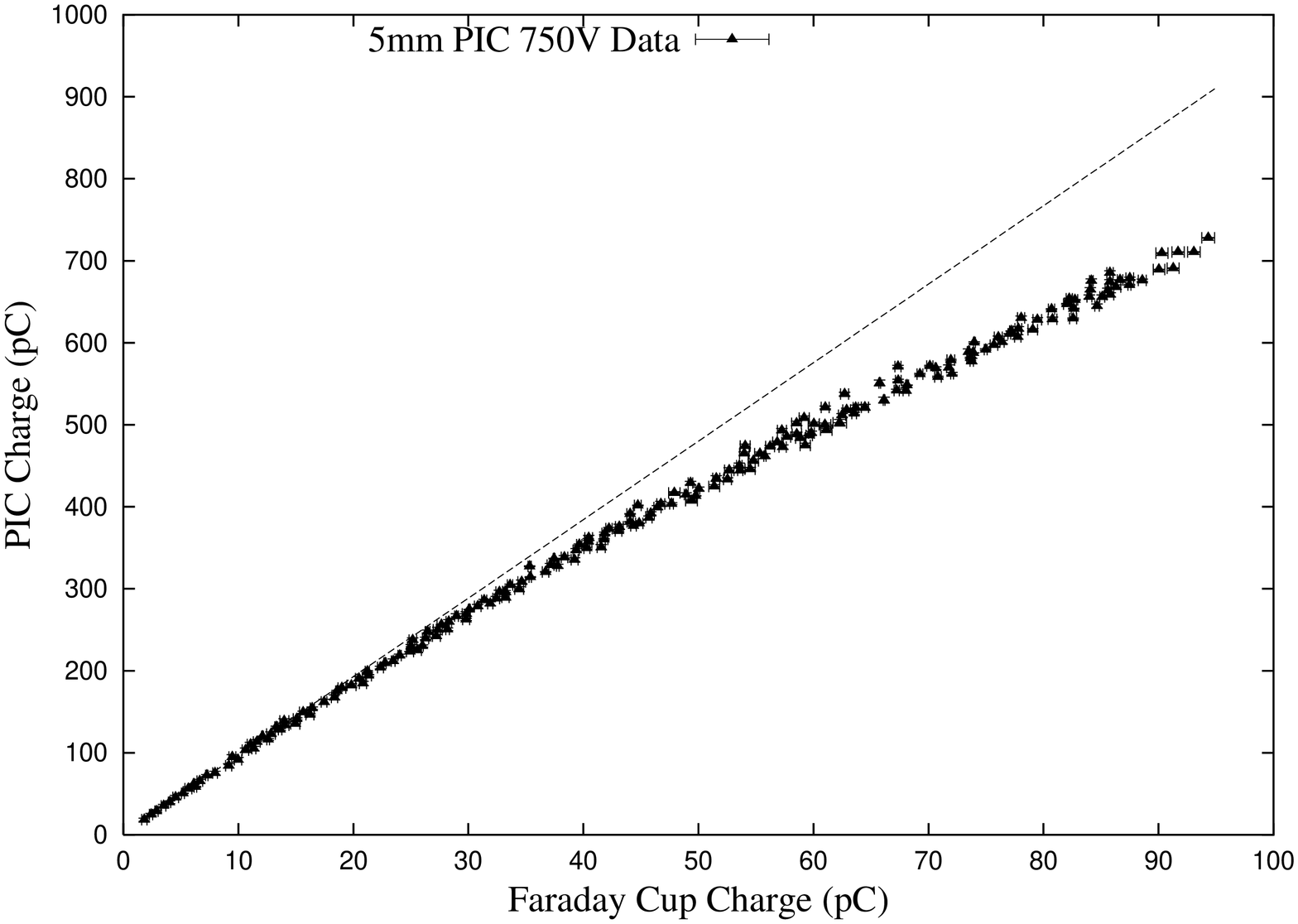}
\end{center}
\caption{Response of the \pic{} versus the \fc{} Charge for 5~mm 750~V data.}
\label{fig:intscan5mm3} 
\end{figure}

\begin{figure} 
\begin{center}

  \includegraphics[width=0.70\linewidth,angle=270]{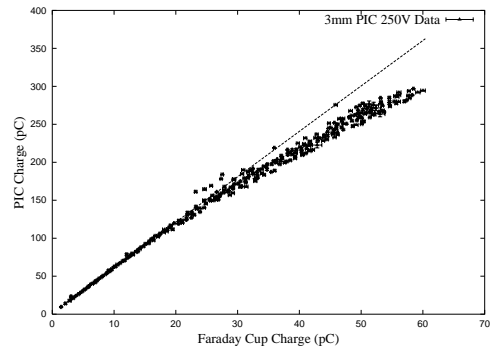}
\end{center}
\caption{Response of the \pic{} versus the \fc{} Charge for 3~mm 250~V data.}
\label{fig:intscan3mm} 
\end{figure}

The linearity measurements for the 5~mm \pic{} at 250, 500 and 750 V
are shown in
Figs.~\ref{fig:intscan5mm},~\ref{fig:intscan5mm2}~and~\ref{fig:intscan5mm3}.
The linearity measurements for the 3~mm \pic{} were performed at 250
and 400~V.  The 3~mm 250~V data are shown in Fig~\ref{fig:intscan3mm};
the 400~V data are similar.  The results of the fits and fit ranges
are summarized in Table~\ref{tab:fit}.  All data are shown with error
bars giving the statistical uncertainty in each bin and the 0.1~pC
SWIC charge transfer error added in quadrature.

The fitting procedure~\cite{numrec} used the values of the \fc{} and
the \pic{} charge with statistical and systematic errors.  For a
linear fit, the data were binned in \fc{} and \pic{} charge.

An important input to the fit is the choice of fit cutoff.  Since the
data are expected to deviate from a linear model at charges beyond the
saturation point, the $\chi^2$/DOF was used to determine the optimal
fit cutoff.  We minimized the $\chi^2$/DOF as a function of the fit
cutoff.  
The fit cutoff, chosen in this way, is one objective measure of the
deviation from linear behavior.

\begin{table*} 
\caption {Fit parameters from linear fits to 5~mm and 3~mm gap helium data. }
\label{tab:fit} 
\begin{tabular}{|l|l|l|l|l|} \hline
 & Slope &  \pic{} inter- & & Range \\ 
Configuration & (pC/pC) & cept (pC) & $\chi^2/$DOF & (pC) \\ \hline
5 mm He-250V & 9.95$\pm$0.22 & -0.80$\pm$1.15 & 17.3/18 & 0-8 \\ \hline
5 mm He-500V &  9.77$\pm$0.09 &  0.22$\pm$0.76 & 30.8/32 & 0-13\\ \hline
5 mm He-750V &  9.57$\pm$0.26 & 1.39$\pm$1.62 & 15.1/15 & 0-11\\\hline
3 mm He-250V &  5.98$\pm$0.03 & 1.51$\pm$0.32 & 58.6/52 & 0-18\\\hline
3 mm He-400V &  5.82$\pm$0.03 & 2.05$\pm$0.30 & 48.4/42  & 0-19\\\hline

\end{tabular} 
\end{table*}

All data in Fig.~\ref{fig:intscan5mm}-\ref{fig:intscan3mm} are clearly
linear below the fit cutoff points, and saturation starts above this
cutoff.  All 5~mm ionization yields (ion pairs/cm) in
Table~\ref{tab:fit} are consistent at the 1.1~$\sigma$ level.  The fit
$y$-intercept is consistent with zero.  The weighted average ratio of
the 3~mm and 5~mm ionization yields is 0.60$\pm$0.01, in excellent
agreement with the expectation of 3/5.

\subsection{Voltage Dependence}

A voltage plateau demonstrates the stable operating range of an
ionization chamber.  For this measurement, the ATF beam was held at
approximately constant intensity while the ionization chamber voltage
was varied from a few volts up to 1000 V.

Using the selection criteria described above, the average \pic{} to
\fc{} yield and RMS were determined from the data.  The beam current
at lower intensities was stable, but at higher intensities varied by
$\approx$20\%.

\begin{figure} 
\begin{center}
\includegraphics[width=0.68\linewidth,angle=270]{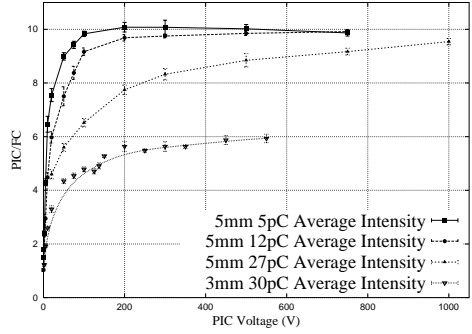}
\end{center}
\caption{\pic{} response as a function of the gap high voltage for
  different nominal intensities for 5~mm and 3mm gap \pic{}.  The
  lines are not a fit to the data points.}
\label{fig:plat5mm} 
\end{figure}

Plateau curves for the 5~mm and 3~mm gap \pic{} in helium gas are
shown in Fig.~\ref{fig:plat5mm}.  The four curves correspond to beam
intensities of 5.7$\pm$0.4, 11.9$\pm$1.7, 26.8$\pm$2.4~pC for
the 5~mm and 30.2$\pm$2.2~pC for the 3~mm PIC.  At the lowest intensity
the ratio \pic{}/\fc{} rises very quickly with voltage and remains
constant beyond $\sim$ 100~V.  At the highest intensity \pic{}/\fc{}
continues to rise even at 1000~V indicating saturation.

\subsection{Saturation}

\begin{figure} 
\begin{center}
\includegraphics[width=0.68\linewidth,angle=270]{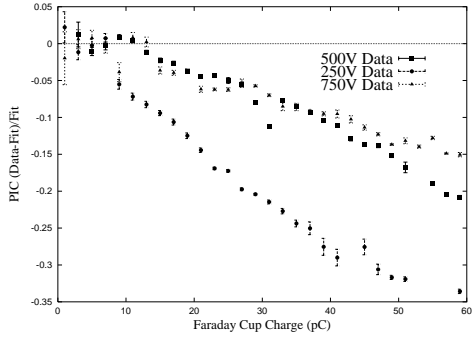}
\end{center}
\caption{\pic{} charge collection efficiency versus integrated beam 
charge measured in the Faraday cup for the 5~mm data.}
\label{fig:eff} 
\end{figure}

To quantify saturation we plot the total collection efficiency which
is defined as the data minus the fit divided by the fit. We then
characterize saturation as a percent deviation from zero.
Fig.~\ref{fig:eff} shows collection efficiency versus \fc{} charge
measured at 250, 500 and 750 V for 5~mm data.
Table~\ref{tab:satlimits} shows the saturation points as determined
from the $\chi^2$ fit (the fit cutoff defined earlier divided by the
area) and 5\% and 10\% saturation points for the 3~mm and 5~mm data.
We found the fit cutoff to be a less sensitive measure of the
saturation point.  The data at higher voltages saturate at higher
intensities, as expected.

The ion chamber, where one pad is held at \mbox{$V(z=0)=0$} and the
other pad at \mbox{$V(z=d)=V_0$}, is analogous to the parallel plate
capacitor.  When a beam of electrons passes through the gas, this
creates a uniform charge distribution, $\rho$.  The electric field,
$E$, in this case is related to the charge distribution by the Poisson
equation, $dE/dz = \rho/\epsilon_0$.  Integrating and applying the
voltage boundary condition, the electric field is given by,
\[ E(z) = \frac{\rho}{\epsilon_0}\left(z - \frac{d}{2}\right) -
\frac{V_0}{d}. \] %
Saturation in this naive model occurs when \mbox{$E(z=d)=0$}, which
corresponds to a critical charge density \mbox{$\rho_c=2\epsilon_0
  V/d^2$} or using the observed ionization response yield of 19.5 ion
pairs/cm and $d=0.5$~cm, \mbox{$\rho_c = 0.708\times V$ pC/cm$^3$},
with V in volts.

\begin{figure} 
\includegraphics[width=0.80\linewidth]{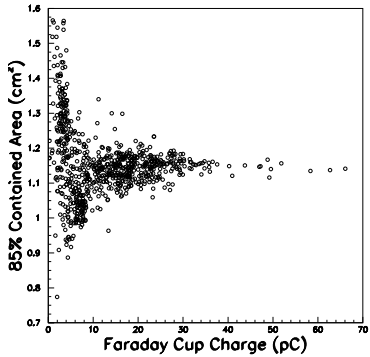}
\caption{The 85\% contained area plots show the area as a function
  of intensity for 5~mm 250~V data.}
  \label{fig:area} 
\end{figure}

To calculate the intensity per unit area the area of the beam spot on
the chamber was measured using three different methods.  For each
method, we used the \bc{} data averaged in the linear region, which
always extended to intensities larger (due to the smaller gap of the
\bc{}) than the saturation point of the \pic{}. The beam spot was
observed periodically (for every fifth pulse) with a fluorescent
screen viewed with a CCD camera and found to be approximately
described by an ellipse.  The area given by,
\[ A =\pi \sigma_{X_c}  \sigma_{Y_c}, \]
with major and minor axes $2\sigma_{X_c}$ and $2\sigma_{Y_c}$ was
determined by fitting a gaussian to the \bc{} profile for the 85\%
contained beam area.  Fig.~\ref{fig:area} shows that the measurement
of the the area was stable and did not degrade as a function of
intensity.

Two additional methods made use of the segmentation of the \bc{}.  A
grid of values was formed by the outer product of the two vectors
consisting of the charge measured in the horizontal and vertical
strips.  The resulting matrix was then normalized by the
total charge.  The first method, (``Box I''), simply added the number
of grid elements with more than 5\% of the maximum charge.  The
second, (``Box II'') summed each grid area weighted by its normalized
charge.

The results are shown in Table~\ref{tab:area}. The three methods agree
within 20\% for all measurements.  The average of these three
measurements is taken to be our area estimate.  The errors on the
measurements are highly correlated and the measurements themselves are
consistent.  Thus, the error on the average was chosen to be the
largest percent error from the individual measurements.  Due to the
large size of the \bc{} strips compared to the beam size, both the
area and error are likely overestimated.

\begin{table*} 
\caption {Area estimates for 5~mm and 3~mm data. }
\label{tab:area} 
\begin{tabular}{|l|l|l|l|} \hline
Data & \multicolumn{3}{c|}{Method (cm$^2$)} \\\hline
& $\pi\sigma_{X_c}\sigma_{Y_c}$ & Box I & Box II \\
& (85\% Containment) & & \\\hline
5 mm 250~V  & 1.16$\pm$0.14 & 1.07$\pm$0.14  & 1.15$\pm$0.10\\\hline
5 mm 500~V  & 1.39$\pm$0.13  & 1.55$\pm$0.21 &  1.08$\pm$0.15\\\hline
5 mm 750~V & 1.16 $\pm$ 0.17 & 1.26 $\pm$ 0.25& 1.12 $\pm$ 0.13\\\hline\hline
3 mm 250~V & 0.99$\pm$0.25 & 1.34$\pm$0.22 & 0.91 $\pm$ 0.21 \\\hline\hline
3 mm 400~V & 0.88$\pm$0.11 & 1.26$\pm$0.16 & 0.81 $\pm$ 0.15\\\hline\hline
\end{tabular} 
\end{table*}

\begin{figure} 
\begin{center}
\includegraphics[width=0.68\linewidth,angle=270]{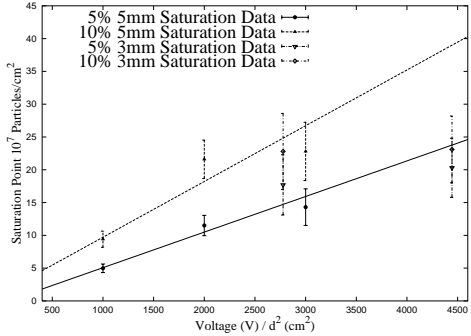}
\end{center}
\caption{Saturation point as a function of the voltage.  Data for 
  the 5\% and 10\% saturation points are shown.  The two linear fits
  include only the 5~mm data.}
\label{fig:vd2} 
\end{figure}

Fig.~\ref{fig:vd2} shows the intensity at which the data saturate as
a function of $V/d^2$ for the 5\% and 10\% level.  The linear fits
shown in the figure include only the 5 mm data.  The 3 mm data are in
good agreement with the fit except for the 400 V 10\% data point which
is lower by 2.5 sigma.  If the 3 mm data are included in the fits, the
fit to the 5 \% data points does not change substantially, but the
$\chi^2/DOF$ for the 10 \% fit worsens from 1.1  to 2.1.
The 5~mm data demonstrate the naive model expectation that the
saturation point behaves linearly in $V/d^2$.  For the 5~mm 500~V
data, the critical charge density corresponds to $11.3 \times
10^7$particles/cm$^2$, which is in good agreement with the measured
value of saturation 11$\pm$ 2 $\times 10^7$ particles/cm$^2$ for the
5\% saturation point.  We note that we have not attempted to
understand the effects of multiplication on ionization yield as well
as on the apparent saturation point.  At high electric fields and
intensities effects of multiplication could make the interpretation of
the data more complex.

\begin{table*} 
\caption {Saturation limits for 5~mm and 3~mm data. }
\label{tab:satlimits} 
\begin{tabular}{|l|l|l|l|} \hline
& \multicolumn{3}{c|}{Saturation Measure} \\ \hline
& $\chi^2$ Fit & 5\% Point & 10\% Point \\ \hline
 & \multicolumn{3}{c|}{$10^7$ particles/cm$^2$} \\ \cline{1-4}
5 mm 250~V Data & 4.4 $\pm$ 0.6 & 5.0 $\pm$ 0.7 & 9.4 $\pm$ 1.2 \\\hline
5 mm 500~V Data & 6.0 $\pm$ 0.8 & 11 $\pm$ 2 &  22 $\pm$ 3 \\\hline
5 mm 750~V Data &  5.8$\pm$1.0 &  14 $\pm$ 3 &  23$\pm$ 4 \\ \hline
3 mm 250~V Data & 10 $\pm$3 & 18 $\pm$ 5 & 23 $\pm$ 6 \\\hline
3 mm 400~V Data & 12 $\pm$3 & 20$\pm$ 4 & 23 $\pm$ 5 \\\hline

\end{tabular} 
\end{table*}

\section{Conclusions}

Results from the ATF measurements have demonstrated that the PIC device
with a 5~mm or smaller gap spacing is suitable for operation in the
high-rate environment expected at the muon monitors of the NuMI beam.
The 5~mm PIC operated linearly with stable response over a voltage
range 250-750~V in helium gas with intensities up to 14 (23) $\times
10^7$ particles/cm$^2$ for a 5\% (10\%) saturation for a short-pulse
beam.  It is expected that for longer duration beam spills of order
10~$\mu$s, the saturation point should increase and this result will
be a lower bound.  Furthermore, the results show that saturation can
be modified in a controlled manner by increasing or decreasing the
externally applied field.

\begin{ack} 
  
  We wish to acknowledge the support of A.~Marchionni, D.~Harris,
  S.~E.~Kopp, R.~M.~Zwaska from the MINOS beam monitoring group for
  useful discussions and helpful suggestions.  We are indebted to
  A.~Franck, D.~Baddorf, B.~Hendricks and J.~Yu for providing the SWIC
  electronics, software and critical expertise.  We thank the ATF
  facility operators as well as Ilan Ben-Zvi and X.~J.~Wang, for
  carefully controlling the beam and actively supporting our tests.
\end{ack}

\bibliography{nim}

\begin{thebibliography}{02}
  
  
\bibitem{NuMiTRD} The MINOS Collaboration, ``Neutrino Oscillation
  Physics at Fermilab:  The NuMI-MINOS Project,'' Fermilab Report
  No. NuMI-L-375 (1998).
  
  
  
\bibitem{atf} T. Tsang \etal, ``Electro-optical measurements of
  picosecond bunch length of a 45 MeV electron beam'', Journal of
  Applied Physics {\bf 89}, 4921-4926, 1 May 2001.

\bibitem{CERN8912} H.~Schonbacher and M.~Tavlet,``Compilation of
  radiation damage test data, pt.1,'' CERN Yellow Report 89-12, 1989.

\bibitem{burrbrown} Burr-Brown Corp., 6730 S. Tucson Blvd, Tucson, AZ 85706. 


\bibitem{Palestini:1998an} S.~Palestini {\it et al.}, Nucl.\ Instrum.\ 
  Meth.\ A {\bf 421}, 75 (1999).
\bibitem{boag} J.~W.~Boag and T.~Wilson, British Journal of Applied
  Physics, {\bf 3}, 222 (1952).
  
\bibitem{swic} \mbox{W.~Kissel, B.~Lublinsky} and A.~Franck, ``New SWIC
  Scanner/Controller System,'' {\it 1995 International Conference on
    Accelerator and Large Experimental Physics Control Systems},
  (1996).
  
\bibitem{numrec} \mbox{W.~H.~Press}, S.~A.~Teukolsky,W.~T.~Vetterling and
  B.~P.~Flannery, ``Numerical Recipies in C,'' Cambridge University
  Press, New~York, 1992.

\end{thebibliography}

\end{document}